\documentclass[twocolumn,preprintnumbers,amsmath,amssymb]{revtex4}
\usepackage{graphicx}
\usepackage{dcolumn}
\usepackage{bm}
\begin{document}
\title{ Negative Group Velocity and Spin-Flip in Microwave Adaptors}

\author{A. Car\^ot,$^1$ H. Aichmann,$^2$,
 and G. Nimtz$^3$\\
$^1$ Anhalt University of Applied Sciences, Bernburger Str. 55, 06366 K\"othen\\
$^2$ Agilent Technologies,Campus Kronberg 7, 61476 Kronberg\\
$^3$ Physics Department, University of Cologne, Z\"ulpicher Str.
77, 50937 K\"oln} \maketitle

\noindent \textbf{Abstract} A Fabry-Perot like interferometer with
two microwaveguide adaptors as reflectors creates a passive
dielectric medium with a negative group delay time due to
polarization shift. A rotational strain of the
polarization vector by one of the adaptors is coupled with a
drastic negative group velocity. The adapted rectangular and
circular waveguides have the same dispersion. The input
rectangular waveguide mode is linearly polarized, whereas the
basic mode of the adapted circular waveguide is circularly
polarized. A 667 wavelengths long circular waveguide connects the
input with the output adaptor. Experiments are performed in the
frequency and in the time domain. We describe, how the helical
polarization change and the spin-flip of the two different
circular wave modes  produce the observed negative group velocity.


\vspace{1cm}

PACS numbers 42.25.-p; 42.25.Ja; 42.25.Gy; 42.50.Ct

\vspace{1cm}
\begin{figure}
\centerline{\includegraphics[width=7cm]{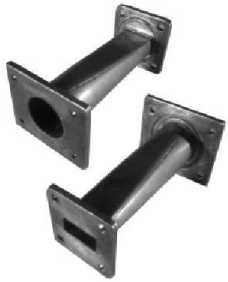}} \caption{
Adapters from rectangular (frequency X-band 8.2 GHz - 12.4 GHz;
wavelength 36.6 mm - 24.2 mm) to circular waveguide and vice
versa. The rectangular and circular guides have the same cut-off
frequency of 6.5 GHz and the same dispersion relation. The inside
guide dimensions are 10.16 mm $\cdot$ 22.86 mm and 27 mm diameter,
respectively. The total adaptor length is 105 mm.} \label{Adaptor}
\end{figure}

\section{Introduction}\label{sec:introduction}
Previously the observation of backward pulse propagation through a
medium with negative group velocity was published by Gehring et
al.~\cite{Geh}. The experiment was carried out in the infrared and
the active medium was designed with the help of an Erbium doped
optical fiber amplifier. Backward waves of guided microwaves have
been studied by Pincherle~\cite{Pin} in 1944, by
Clarricoats~\cite{Clar} in 1963, and different other authors.
Resonant absorbing and amplifying media have been studied in order
to demonstrate negative group velocities of electromagnetic waves
by Segard and Macke~\cite{Seg} and by Wang et al.~\cite{Wang}. In
the previous studies negative group velocity has been observed in
frequency regions near absorption or gain features.  A review
article on controlling light by active media are presented in
Ref.~\cite{Geh2}. Velocity studies on faster than light tunneling
are reviewed in Ref.~\cite{Nimtz4} recently.

Incidentally, a sophisticated mechanical experiment carried out by
Beth provided quantitative evidence of angular photon momentum,
i.e. of the spin in 1936 \cite{Beth}. The photon spin is related
to the electric helical polarization of a photon. This connection
was applied to invert the spins of all the photons in a circularly
polarized light beam by 2$\hbar$ from -1$\hbar$ to +1$\hbar$. Here
we report on a microwave experiment with the passive waveguide
mode adaptors, which can cause an extreme negative group velocity
and a spin-flip. Two adaptors, which are separated by a circular
metal pipe act similarly to a Fabry-Perot interferometer (F-P)
with a negative group delay time at periodical frequency
intervals. However, this phenomenon happens only if the input
polarization of the first and the output polarization of the
second adaptor are not parallel. The microwaveguide adaptors are
displayed in Fig.~\ref{Adaptor}, the angle between the orientation
of the rectangular input and output parts is defined as $\alpha$.
Only for angles $\alpha$ $>$ 0$^0$ and $<$ 180$^0$ the F-P like
behavior and the negative group velocity are observed. In the
experiment, the connecting circular waveguide was turned up
between 0$^0$ and  90$^0$. Polarization is conserved over a
distance of 20 m in the experimental set-up. In the case of the TV
satellite communication, polarization is conserved over a distance
of 35,786 km.

A F-P interferometer can be described as a one-dimensional cavity
constructed by two tunneling barriers. In such Fabry-Perot
set-ups, superluminal velocities have been observed with microwave
 \cite{Nimtz1,Nimtz2} and infrared digital signals
\cite{Longhi}. Negative group velocities in special Fabry-Perot
structures have been investigated in Refs.\cite{Piere,Ao}, for
instance.
\begin{figure}
\centerline{\includegraphics[width=7cm]{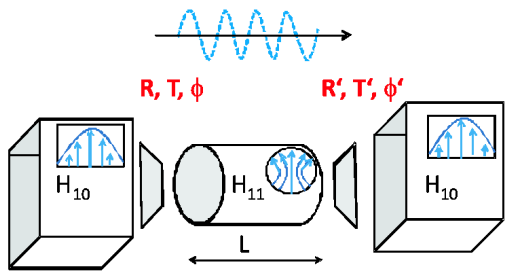}}
\caption{Experimental set-up. From left: Signal input via
rectangular wave guide with H$_{10}$ mode adapter to circular wave
guide with H$_{11}$ mode (R, T, $\Phi$). The second adaptor
transmits the circular mode back to a rectangular wave guide with
H$_{10}$ mode (R', T', $\Phi$'). R, R', $\Phi$, $\Phi$' L are
reflections, phase shifts at the adaptors, and L the length of the
circular waveguide between the adaptors. The field distribution in
the waveguide cross section is sketched.} \label{setup}
\end{figure}
In the investigated set-up Fig.~\ref{setup} periodical
interferometer structures of the complex transmission and
reflection begin above 0$^o$ of the angle $\alpha$ and increases
up to 90$^o$.

The frequency band width between two resonance transitions of a
F-P interferometer is given by the relation

\begin{eqnarray}
\Delta \nu & = & \frac{c}{n 2 L}, \label{Frequenz}
\end{eqnarray}
where c is the velocity of light in vacuum, n is the refractive
index of the material between the two reflectors, and L the
distance of the reflectors. This relation is fulfilled in the
studied experimental design with mode adaptors as mirrors. For
instance, with L = 20 m and n $\approx$ 1.5  we obtain a value of
$\approx$ 5 MHz.

As we shall see below, only the frequency band width between the
transmission maxima of the adaptor design agrees with a F-P
interferometer but not the shape of the transmission spectra or
disappearing periodical structures as $\alpha$ goes to zero. The
resonance dips are maximal, when the input and output rectangular
wave guides are oriented perpendicularly. The transmission of the
set-up is maximal, whereas the periodic structures disappear at an
angle $\alpha$ = 0. Parallel orientation of the input and output
rectangular waveguides represents the normal technical use of such
a set-up.

\section{Experimental Results}

The rectangular X-band waveguide and the circular guide have the
same frequency dependent dispersion and the same cut-off frequency
of 6.5 GHz. The set-up is sketched in Fig.~\ref{setup}. The input
in the rectangular waveguide with the H$_{10}$ mode excites
circular H$_{11}$ modes in the first adapter. The electric field
distribution of the two modes are sketched in Fig.\ref{setup}. The
circular modes (right and left circularly polarized) are
transmitted on a circular wave guide of length L and transduced at
the second adapter back to the H$_{10}$ mode in the rectangular
waveguide. Depending on the angle between the rectangular input
and output waveguide we obtain weak or strong F-P resonances.

The phase, transmission, and reflection structures are periodical
with a frequency band width according to the Eq.~\ref{Frequenz}.
We have measured lengths L of 0.2, 5, and 20 m. The transmission
shows a drastic deviation from a classical F-P interferometer
since the maxima are very broad, whereas the minima are very
narrow in frequency.

\begin{figure}
\centerline{\includegraphics[width=9cm]{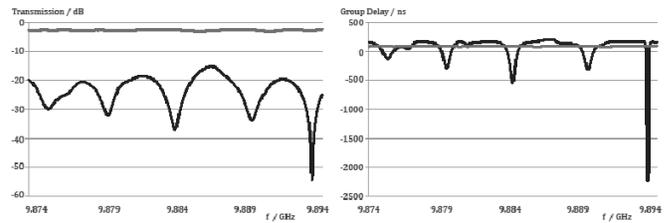}}
\caption{Transmission vs frequency of the 20 m circular waveguide.
The adapters are oriented $\alpha$ = 0 (attenuation of $\approx$
-2.5 dB, grey line) and $\alpha$ = 90$^0$ (attenuation
oscillations between -15 dB and -55 dB). The figure displays the
frequency range between 9.874 GHz and 9.894 GHz. Average
periodicity of the structure is $\approx$ 5 MHz. Right: Group
delay time vs frequency of the same set-up. (Remember the vacuum
delay of 20 m is 67 ns, whereas the transmission time for $\alpha$
= 0$^0$ is $\approx$ 100 ns, grey line). The negative delay time
of 2.2 $\mu$s equals that measured in the time domain also at
9.865 GHz as seen in Fig.~\ref{Adap3}.} \label{Trans1}
\end{figure}

At resonance frequencies, transmission minima up to -55 dB were
measured. The important result at the transmission minima was a
negative phase shift and thus a negative group delay time up to
-2.2 $\mu$s.

The measurements were carried out in the frequency domain with a
network analyzer (Rhode-Schwartz ZVK). In the time domain
measurements have been done with an oscilloscope (HP-Infinium
2GSa/s/4ch). The time domain results are in agreement with those
obtained in the frequency domain, time domain examples are
presented in Fig.~\ref{Adap3}.

The frequency range of the Ku-band is 12.4 GHz - 18 GHz. The
cut-off frequency of the circular and the rectangular guides is
9.5 GHz. The inside waveguide dimensions are 15.8 mm $\cdot$ 7.9
mm  the corresponding circular waveguide has an inside diameter of
18.5 mm. For comparison we have investigated the same quantities
with Ku-band adaptors and L = 0.2 m. The results are in agreement
with those observed in the X-band.

\section{Discussion}
\begin{figure}
\centerline{\includegraphics[width=9cm]{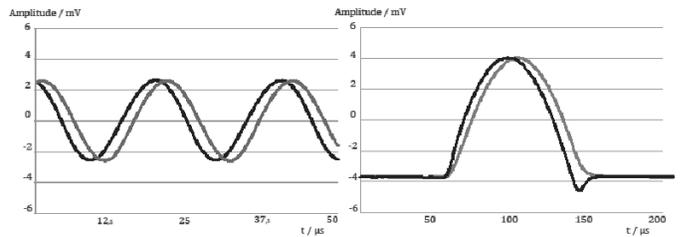}} \caption{Left:
Delay time of a  polarization turned AM 48 kHz wave at a carrier
frequency of 9.893 GHz. This fast normalized  wave is compared
with the grey wave measured at the input of the 20 m long device.
The negative shift of $\approx$ 2 $\mu$s points to a negative
group velocity of 0.03 c. For instance, the maxima of the fast
wave leaves the 20 m waveguide before it has entered it. Remember
this happens in a passive medium. Right: Transmission vs time of a
signal output at the 20 m circular waveguide (carrier frequency
9.893 GHz). The normalized fast transmitted signal is compared
with the grey input signal. The signal output has similar waveform
as the infrared one obtained in the active medium displayed in
Ref.\cite{Geh}.} \label{Adap3}
\end{figure}

The experimental results, which are displayed in
Figs.~\ref{Trans1}-\ref{Adap3} are obtained for an adapter
distance L = 20 m, which corresponds to 666 wavelengths. Actually,
the right signal in Fig.~\ref{Adap3} has the same reshaped
structure as the infrared backward wave in the active media
observed in Ref.~\cite{Geh}. Depending on the angle between input
and output polarization, a negative group delay and thus negative
group velocity was observed. Examples with a negative velocity of
the order of -0.03 c are presented in Figs. \ref{Adap3}. The
velocity is calculated according to the phase time relations
Eq.~\ref{Freq} and from the time shift $\Delta$t $\approx$ 2.2~
$\mu$s by Eq.~\ref{time}, in agreement with the time domain data
of Fig.~\ref{Adap3}.

The mode and thus the state vector of the photons  are given by
the first linearly polarized H$_{10}$ mode, which is transduced
into right and left circularly polarized H$_{11}$ modes in the
first adaptor. If the input and the output polarization of the
rectangular waveguides are equal, the linearly polarized H$_{10}$
mode has a small attenuation of -2.5 dB, Fig.~\ref{Trans1}. The
attenuation is essentially due to waveguide losses. With
increasing angle $\alpha$, the transmission decreases since
reflection takes place at twisted rectangular output waveguide.
For $\alpha$ = 90$^o$ we have a reduced transmission to $\leq$ -15
dB in the X-band set-up with L = 20 m.

There are several theoretical approaches to explain a negative
group delay time of special F-P interferometers. For example, the
F-P like behavior of waveguide discontinuities by tapered
waveguides with the same mode are studied in Ref.\cite{Ao}. The
group delay is given  by the relation

\begin{eqnarray}
\tau_g  & = & \frac{d\varphi}{d\omega}\\
v_g & = & L/\tau_g, \label{Freq}\\
v_g & = & \frac{L}{\Delta t + L/c}, \label{time}
\end{eqnarray}
where   $\tau_g$ and $v_g$ are the group delay time and the group
velocity respectively. $\varphi$ and $\omega$ are phase and
angular frequency of the wave. $\Delta$t is the measured time
shift compared with the time spent traversing the same vacuum
distance L.

\begin{equation}
\tau_g = \left( \frac{1 + R'}{1 - R}\right) \frac{L}{v_g} +
\left(2\frac{d\phi_t}{d\omega} - \frac{2 R'}{1 + R'}
\frac{d\phi'_r}{d\omega}\right),
\end{equation}
where R, R', $\phi_t$, $\phi_r$, L, and $\omega$ are reflections,
phase shifts at the adaptors, L the length of the circular
waveguide between the adaptors, and $\omega$ the angular
frequency.

In the case that the last component of the delay time equation
dominates, the group velocity becomes negative;

\begin{equation}
\left(\frac{2 R'}{1 + R'} \frac{d\phi'_r}{d\omega}\right) > \left(
\frac{1 + R'}{1 - R}\right) \frac{L}{v_g} +
\left(2\frac{d\phi_t}{d\omega}\right)
\end{equation}
This approach seems not to be appropriate to solve the problem in
question,  since the quasi mirrors are given by a waveguide
discontinuity with the same mode. In our experiment a mode
transition is causing the reflection. The latter process is
related to the polarization and thus to the spin state of the
photons.

\begin{figure}
\centerline{\includegraphics[width=7cm]{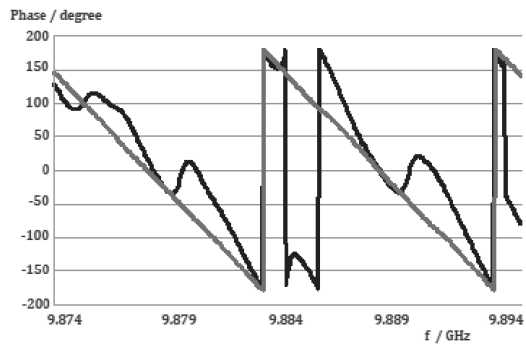}} \caption{Phase
vs frequency showing the frequency regimes with negative phase.
The linear dispersion of the parallel oriented adaptors (grey)
corresponds to a group velocity of 0.68 c and a refractive index
of 1.5.} \label{Group4}
\end{figure}

Ibanescu et al.  studied symmetry breaking axially uniform
waveguides \cite{Iban}. They have calculated anomalous dispersion
relations by symmetry breaking. They deduced that anomalous
dispersion might take place, if two modes of E (TM)   and  H (TE)
symmetry are close near the long wavelength limit. However, this
is not the case for the studied H$_{10}$ and H$_{11}$ modes.

The observation in the pseudo F-P interferometer is described as
follows. The H$_{10}$ input mode is linearly polarized and is
resolved into two equal amplitude circularly polarized waves
H$_{11}$ of opposite hand. In this way the input polarization and
spin is conserved. According to the expressions for the two
circularly modes H$_{11}$ its angle dependence is
\begin{eqnarray}
E_{rr} & = & A J \,\,\Xi\, exp\,[i(\omega t - k z  -\pi/2)]\\
E_{rl} & = & A J\,\,\Xi\, exp\,[i(\omega t -
k z  +\pi/2)]
\end{eqnarray}
where A,J,$\Xi$ are amplitude and Bessel function
terms~\cite{Marc}, and E$_{rr}$, E$_{rl}$ are the radial electric
fields right hand and left hand. k, z are the wave number and the
propagation direction. Both oppositely polarized circular waves synthesize a linearly polarized wave.

Each of the two circular waves, which differ in phase by $\mp$
$\pi$/2 are reflected at the second adaptor by $\pi$. Thus the
total reflected polarization is turned by $\pi$/2 and travels back
to the first adapter and is reflected once more now at the input
adaptor forming a long cavity at the proper multiple half
wavelength. This takes place at this L of 20~m all 5 MHz. Remarkably, the first adaptor becomes reflective by the second
adaptor's reflection. A cavity is formed by the change of the direction of the synthesized linear polarization at the adaptors.

The negative phase
vs frequency regime is shown in Fig.~\ref{Group4}.
At the resonance levels the group delay time becomes superluminal
and actually negative up to some $\mu$s for a small transmission
in a narrow frequency band. The frequency band width varies in the
differently shaped resonance dips. This is due to the frequency dependent
quality factor of the 20 m long metal pipe resonator.

The reflection and transmission at the mode adaptors experience a
negative phase shift at the dispersion around the resonance as
displayed in Fig.~\ref{Group4}. The resonances are small dips in
transmission as in the case of the classical frequency measurement
by tunable slightly coupled cavities. This is opposite to the
Fabry-Perot interferometer, where the transmission is near to 1. A
narrow resonance line has a correspondingly larger negative group
delay time and smaller transmission depending on the quality
factor in agreement with the measurement.

\section{Conclusions}
Summing up we have observed that mode adaptors in non-parallel
orientations exhibit the following properties: The waveguide
adaptors act as reflectors if they are not parallel oriented.
Transmission decreases and negative group velocities occur with an
increasing angle from parallel to perpendicular orientation. The
dispersion periodicity of 5 MHz and negative dispersion regions
are presented for a 667 wavelength long F-P like set-up in
Figs.~\ref{Trans1}, \ref{Group4}. The negative group delay is
observed in the narrow frequency regime of negative dispersion.
The most negative delay time occurs at the turning point of the
negative part of the dispersion function, Fig.~\ref{Group4}. In
consequence of the reflection at the metal wall of the twisted
waveguide the two helical waves change the handiness and perform a
spin-flip see Ref.\cite{Beth} for instance.

\section{Acknowledgements}
We gratefully acknowledge helpful discussions on time reversal and
on the Beth experiment with Paul Bruney and Friedrich Wilhelm
Hehl.


\begin{thebibliography}{99}
\bibitem{Geh} G. M. Gehrig, A. Schweinsberg, C. Barsi, N.
Kostinski, and R. B. Boyd, Science, \textbf{312}, 895 (2006)
\bibitem{Pin} L. Pincherle, Phys. Rev. \textbf{66}, 118 (1944)
\bibitem{Clar} P. J. B. Clarricoats, Proc. I.E.E. \textbf{110}, 261 (1963)
\bibitem{Seg}B. Segard and B. Macke, Phys. Lett. A \textbf{109}, 213
(1985)
\bibitem{Wang} L. Wang, A. Kuzmich, and A. Dogariu, Nature \textbf{406},
277 (2000)
\bibitem{Geh2}R. W. Boyd and D. J. Gauthier, Science, \textbf{326}, 1074
(2009)
\bibitem{Nimtz4}G. Nimtz, Found. Phys. \textbf{41}, 1193 (2011)
\bibitem{Beth}R. A. Beth, Phys. Rev. \textbf{50}, 115 (1936)
\bibitem{Nimtz1}A. Enders and G. Nimtz, Phys. Rev. B \textbf{47}, 9605[1993]
\bibitem{Nimtz2}G. Nimtz, A. Enders, and H. Spieker, Phys. I France \textbf{4},565(1992)
\bibitem{Longhi}S. Longhi,M. Marano, M. Belmonte, and P. Laporta,IEEE J. Selected Topics Quantum Electronics,
\textbf{\textbf{9}},4(2003)
\bibitem{Esposito}S. Esposito, arXiv:quant-ph/0209018 v l 2.Sep(2002)
\bibitem{Nimtz3}G. Nimtz, Found. Phys. \textbf{39}, 1346 (2009)
\bibitem{Hartman}T. Hartman, J. Appl. Phys. \textbf{33}, 3427 (19962)
\bibitem{Piere}P. Tournois, IEEE J. Quantum Electronics \textbf{33}, 519(1997)
\bibitem{Ao}H. Y. Yao and T. H. Chang, Progress in Electromagnetic
Research, \textbf{122}, 1 (2012)
\bibitem{Nimtz5} G. Nimtz, LNP, \textbf{702}, 506[2006)
\bibitem{Iban} M. Ibanescu, S. G. Johnson, D. Roundy, C. Luo, Y.
Fink, and J. D. Joannopoulos, Phys. Rev. Lett. 92, 063903 (2004)
\bibitem{Marc} N. Marcuvitz, Waveguide Handbook, Dover
Publications, Inc., New York (1951)

\end{thebibliography}
\end{document}